\documentclass[12 pt,twoside,aps,prd,amsmath,amssymb,
tightenlines,showpacs,showkeys,eqsecnum]{revtex4}
\usepackage{mathptmx}
\usepackage{bm}
\usepackage{graphicx}
\usepackage{hyperref}
\def \myfigures #1#2#3#4#5#6#7#8
{\begin{figure}[ht]
    \begin{center}
        \includegraphics[width=#2 \textwidth]{#1.eps}
        \hfill
        \includegraphics[width=#6 \textwidth]{#5.eps}
        \parbox[t]{#4\textwidth}{\caption {#3}\label{#1}}
        \hfill
        \parbox[t]{#8\textwidth}{\caption {#7}\label{#5}}
    \end{center}
\end{figure} }

\def\myfigure #1#2#3#4
{\begin{figure}[ht]\begin{center}
\includegraphics[width=#2 \textwidth]{#1.eps}
\parbox[t]{#4\textwidth}{\caption{#3}\label{#1}}
\end{center}\end{figure}}

\date{\today}
\begin{document}
\title{Bianchi type-II cosmological model: some remarks}
\author{Bijan Saha}
\affiliation{Laboratory of Information Technologies\\
Joint Institute for Nuclear Research, Dubna\\
141980 Dubna, Moscow region, Russia} \email{bijan@jinr.ru}
\homepage{http://bijansaha.narod.ru/}

\begin{abstract}
Within the framework of Bianchi type-II (BII) cosmological model the
behavior of matter distribution has been considered. It is shown
that the non-zero off-diagonal component of Einstein tensor implies
some severe restriction on the choice of matter distribution. In
particular for a locally rotationally symmetric Bianchi type-II (LRS
BII) space-time it is proved that the matter distribution should be
strictly isotropic if the corresponding matter field possesses only
non-zero diagonal components of the energy-momentum tensor.
\end{abstract}



\maketitle

\section{Introduction}

Spatially homogeneous and anisotropic cosmological models play a
significant role in the description of large scale behavior of
universe. In search of a realistic picture of the early universe
such models have been widely studied in framework of General
Relativity. In this note we confine our study within the scope of a
Bianchi type-II space-time, which has recently been studied by a
number of authors. Shri Ram and P. Singh (1993) presented the
analytical solutions of the Einstein-Maxwell equations for
cosmological models of LRS Bianchi type-II, VIII and IX. Two-fluid
Bianchi type-II cosmological models were studied by Pant and Oli
(2002). A Bianchi type-II cosmological model with constant
deceleration parameter considered by Singh and Kumar (2003).
Belinchon (2009a, 2009b) studied the massive cosmic string within
the scope of BII model. LRS BII cosmological models in presence of
massive cosmic string and varying cosmological constant were studied
by Pradhan {\it et. al.} (2010), Kumar (2010) and Yadav {\it et.
al.} (2010), respectively. The main aim of this report is to show
that one has to take into account all the 10 Einstein equations in
order to write the correct energy-momentum tensor for the source
field.

\section{The Metric and Field  Equations}
We consider a homogeneous Bianchi type-II space-time with the line
element
\begin{equation}
ds^{2} = -dt^2 + A^2(dx - zdy)^2 + B^2 dy^2 + C^2 dz^2, \label{BII}
\end{equation}
with $A,\,B,\,C$ being the function of $t$ only. The metric
\eqref{BII} possesses the following non-zero components of Einstein
tensor:
\begin{subequations}
\label{ET}
\begin{eqnarray}
G_1^1 &=& \frac{\ddot B}{B} + \frac{\dot C}{C} + \frac{\dot
B}{B}\frac{\dot C}{C} - \frac{3}{4} \frac{A^2}{B^2C^2},
\label{G11}\\
G_2^2 &=& \frac{\ddot C}{C} + \frac{\dot A}{A} + \frac{\dot
C}{C}\frac{\dot A}{A} + \frac{1}{4} \frac{A^2}{B^2C^2},
\label{G22}\\
G_3^3 &=& \frac{\ddot A}{A} + \frac{\dot B}{B} + \frac{\dot
A}{A}\frac{\dot B}{B} + \frac{1}{4} \frac{A^2}{B^2C^2},
\label{G33}\\
G_0^0 &=& \frac{\dot A}{A}\frac{\dot B}{B} + \frac{\dot
B}{B}\frac{\dot C}{C} + \frac{\dot C}{C}\frac{\dot A}{A} -
\frac{1}{4} \frac{A^2}{B^2C^2}, \label{G00}\\
G_2^1 &=&  z \Biggl[\frac{\ddot A}{A} - \frac{\ddot B}{B} +
\frac{\dot C}{C}\frac{\dot A}{A} - \frac{\dot B}{B}\frac{\dot C}{C}
+ \frac{A^2}{B^2C^2}\Biggr]. \label{G12}
\end{eqnarray}
\end{subequations}
Here over-dot denotes derivation with respect to time.  It can be
easily verified that
\begin{equation}
G_2^1 = z \Bigl[G_2^2 - G_1^1\Bigr]. \label{G12n}
\end{equation}

Let us now note that the system
\begin{equation}
G_\nu^\mu = R_\nu^\mu - \frac{1}{2} R \delta_\nu^\mu = \kappa
T_\nu^\mu, \label{EEg}
\end{equation}
comprises of 10 equations. Depending on the concrete metric and
source field there might a fewer number of equations indeed. But one
has to remember that any equation of the system \eqref{EEg} can be
ignored if and only if

(i) both the left and right hand sides of the equation in question
are identically zero;

(ii) this equation can be obtained from other equations after some
manipulations.

For example, the equation
\begin{equation}
G_2^1 = \kappa T_2^1, \label{EEg12}
\end{equation}
can be ignored only if the equality
\begin{equation}
T_2^1 = z \Bigl[T_2^2 - T_1^1\Bigr] \label{T12n}
\end{equation}
holds.

In all other cases one has to write down all the equations. Those
equations give relations between the metric functions (if $G_j^i \ne
0$ and $T_j^i \equiv 0$, as in case of a BVI space-time with only
non-zero diagonal components of the energy-momentum tensor [cf.
e.g., Saha 2004]) or material field functions (if $G_j^i \equiv 0$
and $T_j^i \ne 0$, as in case of a BI space-time with
electro-magnetic fields with $A_\mu = (0,\,A_1,\,A_2,\,A_3)$  [cf.
e.g., Rybakov {\it et. al.} 2010a, 2010b]).

Let us now study the BII cosmological model for a source field  with
the non-zero diagonal components only, i.e., with the
energy-momentum tensor given by
\begin{equation}
T_\nu^\mu = {\rm diag}\Bigl[T_0^0,\, T_1^1,\,T_2^2,\,T_3^3\Bigr],
\label{emt}
\end{equation}

Let us now write the corresponding Einstein field equations:
\begin{subequations}
\label{EE}
\begin{eqnarray}
\frac{\ddot B}{B} + \frac{\dot C}{C} + \frac{\dot B}{B}\frac{\dot
C}{C} - \frac{3}{4} \frac{A^2}{B^2C^2} &=& \kappa T_1^1,
\label{EE11}\\
\frac{\ddot C}{C} + \frac{\dot A}{A} + \frac{\dot C}{C}\frac{\dot
A}{A} + \frac{1}{4} \frac{A^2}{B^2C^2} &=& \kappa T_2^2,
\label{EE22}\\
\frac{\ddot A}{A} + \frac{\dot B}{B} + \frac{\dot A}{A}\frac{\dot
B}{B} + \frac{1}{4} \frac{A^2}{B^2C^2} &=& \kappa T_3^3,
\label{EE33}\\
\frac{\dot A}{A}\frac{\dot B}{B} + \frac{\dot B}{B}\frac{\dot C}{C}
+ \frac{\dot C}{C}\frac{\dot A}{A} - \frac{1}{4} \frac{A^2}{B^2C^2}
&=& \kappa T_0^0, \label{EE00}\\
\frac{\dot A}{A} - \frac{\dot B}{B} + \frac{\dot C}{C}\frac{\dot
A}{A} - \frac{\dot B}{B}\frac{\dot C}{C} + \frac{A^2}{B^2C^2} &=& 0.
\label{EE12}
\end{eqnarray}
\end{subequations}
From \eqref{G12n} follows the certain relation between the
components of the energy-momentum tensor, namely $T_1^1 = T_2^2$.

Hence we can conclude that Bianchi type-II cosmological model given
by the metric element \eqref{BII} does not allow the situation when
$T_1^1 \ne T_2^2$, i.e., in this case the cosmic string cannot be
directed along either $x$ or $y$ axes. Moreover, in the case when
$T_1^1 = T_2^2$ one needs not to take into account the Eqn.
\eqref{EE12}, as well as in case of isotropic distribution of matter
when $T_1^1 = T_2^2 = T_3^3$. It should be emphasized that in case
of a LRS BII model when we have $A = C$ in \eqref{BII}, the equality
$G_1^1 = G_3^3$ implies $T_1^1 = T_3^3$ and \eqref{T12n} leads to
$T_1^1 = T_2^2$ if the matter distribution is given by \eqref{emt}.
It means that a LRS BII model filled with matter field given by
\eqref{T12n} does not allow anisotropic distribution of matter.

This simple example shows that one should be very careful in
choosing source fields for the cosmological models.

\section{conclusion}
Within the scope of BII cosmological model it is shown that the
non-zero off-diagonal component of the Einstein tensor implies
severe restriction of the components of energy-momentum tensor. In
case of LRS BII space-time the model allows only isotropic
distribution of source field.



\end{document}